\documentclass[pdflatex,sn-mathphys-num]{sn-jnl}% Math and Physical Sciences Numbered Reference Style 
\usepackage{graphicx}%
\usepackage{multirow}%
\usepackage{amsmath,amssymb,amsfonts}%
\usepackage{amsthm}%
\usepackage{mathrsfs}%
\usepackage[title]{appendix}%
\usepackage{xcolor}%
\usepackage{textcomp}%
\usepackage{manyfoot}%
\usepackage{booktabs}%
\usepackage{algorithm}%
\usepackage{algorithmicx}%
\usepackage{algpseudocode}%
\usepackage{listings}%

\theoremstyle{thmstyleone}%
\theoremstyle{thmstyletwo}%

\theoremstyle{thmstylethree}%

\begin{document}

\title[Acoustic shape-morphing micromachines]{Acoustic shape-morphing micromachines}

\author*[1]{\fnm{Xiaoyu} \sur{Su}}\email{suxyu@pku.edu.cn}

\affil[1]{\orgdiv{School of Mechanical Engineering}, \orgname{Northwestern Polytechnical University}, \orgaddress{\street{Youyi West Road}, \city{Xi'an}, \postcode{710072}, \state{Shaanxi}, \country{China}}}

\abstract{Shape transformation is crucial for the survival, adaptation, predation, defense, and reproduction of organisms in complex environments. It also serves as a key mechanism for the development of various applications, including soft robotics, biomedical systems, and flexible electronic devices. However, among the various deformation actuation modes, the design of deformable structures, the material response characteristics, and the miniaturization of devices remain challenges. As materials and structures are scaled down to the microscale, their performance becomes strongly correlated with size, leading to significant changes in, or even the failure of, many physical mechanisms that are effective at the macroscale. Additionally, electrostatic forces, surface tension, and viscous forces dominate at the microscale, making it difficult for structures to deform or causing them to fracture easily during deformation. Moreover, despite the prominence of acoustic actuation among various deformation drive modes, it has received limited attention. Here, we introduce an acoustical shape-morphing micromachine (ASM) that provides shape variability through a pair of microbubbles and the micro-hinges connecting them. When excited by external acoustic field, interaction forces are generated between these microbubbles, providing the necessary force and torque for the deformation of the entire micromachine within milliseconds. We established programmable design principles for ASM, enabling the forward and inverse design of acoustic deformation, precise programming, and information storage. Furthermore, we adjusted the amplitude of acoustic excitation to demonstrate the controllable switching of the micromachine among various modes. By showcasing the micro bird, we illustrated the editing of multiple modes, achieving a high degree of controllability, stability, and multifunctionality. Our findings pave the way for new research avenues in acoustic and deformable micromachines, offering novel design paradigms and development opportunities in the fields of robotics, metamaterials, adaptive optics, flexible electronics, and microtechnology.}

\keywords{Acoustic, Wireless micromachines, Shape transformation, Infinite modes, programmable}

\maketitle

\section{Introduction}\label{sec1}

Shaper-morphing mechanisms are critical mechanisms for organisms to survive and adapt to their environments. The fringed-lizard spreads its umbrella-like collar membrane during threat displays and mating rituals\textsuperscript{\cite{peacock2022acoustical}},while the sensitive plant closes its leaves to avoid damage from storms\textsuperscript{\cite{volkov2010mimosa}}. Due to their exceptional environmental sensitivity and adaptability, these critical morphing mechanisms have found widespread applications at the macroscopic scale, including robotic systems\textsuperscript{\cite{sun2023embedded,kotikian2019untethered}}, wearable devices\textsuperscript{\cite{qin2021exoform,fei2021cephalopod}}, medical devices\textsuperscript{\cite{zarek20174d,qu2023smart}}, artificial muscles\textsuperscript{\cite{yang2023precisely,fei2021cephalopod}}, and materials science\textsuperscript{\cite{bastola2021shape,oliver2016morphing}}. For instance, soft robotic systems used for mechanical propulsion\textsuperscript{\cite{liu2023ultrafast,fu2018morphable}}, actuators based on Kirigami mechanisms\textsuperscript{\cite{rafsanjani2018kirigami,tang2019programmable,zhu2022kirigami}}, shape-memory polymers and alloys\textsuperscript{\cite{rodriguez2016shape,ze2020magnetic,dezaki2022adaptive}}, liquid crystal elastomers\textsuperscript{\cite{zhao2022twisting,sun2023additive}} and conductive polymers\textsuperscript{\cite{hajiesmaili2019reconfigurable,aksoy2020reconfigurable}}. However, most existing robots are on the millimeter or centimeter scale, making miniaturization challenging. Additionally, these devices often require tethered connections to external sources\textsuperscript{\cite{ma2013controlled,ma2012design}}, further limiting their applicability at the microscale. Moreover, as static friction and adhesion forces increase, surface forces such as van der Waals and electrostatic forces become predominant at the microscale. Consequently, small-scale structures are more susceptible to maintaining rigidity and experiencing fractures\textsuperscript{\cite{purcell2014life,bazant2005scaling}}, which renders the development of microscale variants more challenging.

With the rapid advancements in smart functional materials\textsuperscript{\cite{sidorenko2007reversible,van2009printed,zeng2014high}}, micro/nano manufacturing technologies\textsuperscript{\cite{schaedler2011ultralight,buckmann2012tailored}}, precise manipulation\textsuperscript{\cite{alogla2015micro,power2018monolithic}}, and multi-physics driving strategies\textsuperscript{\cite{xia2010ferrofluids,chen2017hybrid}}, the performance of morphing machines has significantly improved, leading to advancements toward miniaturization, increased intelligence, and multifunctionality. These small-scale morphing micro-machines have widespread applications in various fields, including biomedical engineering\textsuperscript{\cite{kuo2014hydrogel,yang2019targeted}}, microfluidics\textsuperscript{\cite{yu2001responsive,jamal2011differentially}}, and microactuators\textsuperscript{\cite{kim2020shape,liu2021micrometer}}. Researchers have employed multi-physics field-driven methods to control small, untethered robots, utilizing various stimuli such as chemical gradients\textsuperscript{\cite{hu2020botanical,huang2020four,narupai20214d,dong20204d}}, light\textsuperscript{\cite{cheng2020kirigami,lv2016photocontrol,chen2023light,wang2021light,zhang2022light}}, heat\textsuperscript{\cite{hwang2022shape,aharoni2018universal,ding2017direct,zheng2023electrodeposited}}, electric fields\textsuperscript{\cite{he2019electrically,wang2022cilia,miskin2020electronically,liu2021micrometer,reynolds2022microscopic}}, and magnetic fields\textsuperscript{\cite{hu2021magnetic,zhang20233d,alapan2020reprogrammable,ji2021enhanced}}. While these strategies address the challenges posed by microscale effects, they still present significant limitations. pH-responsive hydrogel materials exhibit limited sensitivity and slow response times, as they can only react within a specific pH range. The choice of solvent can influence the material's behavior and applications, thereby constraining its potential uses\textsuperscript{\cite{chen2021intelligent}}. Shape memory alloys or polymers are commonly used as temperature-responsive deformable materials; however, morphing mechanisms made from these materials exhibit slow response times and face challenges in miniaturization\textsuperscript{\cite{kim2023shape}}. Photoresponsive morphing structures are constrained by environmental factors, and the wavelength range of photoresponsive materials is limited. Additionally, light is prone to scattering and reflection, which restricts the functionality and repeatability of photoresponsive morphing mechanisms\textsuperscript{\cite{yang2019photoresponsive}}. Deformation structures driven by electric and magnetic fields have garnered significant attention\textsuperscript{\cite{bastola2021shape,ahn2022review}}. However, these methods require complex pre-programming to produce micro-machines compatible with such approaches, including pre-designed magnetic moments, multilayer structures, and multi-step processes. Furthermore, when scaling down, the underlying physical mechanisms governing their operation may be significantly affected or may fail.

Ultrasound fields exhibit excellent properties such as high penetration, flexibility, low cost, and biocompatibility, and can generate significant power. Despite the exciting prospects that acoustics offer for transformable micro-machines, this area has received minimal attention\textsuperscript{\cite{zhang2024sonotransformers}}. Acoustics provides an effective method for generating controllable deformation in micro-machines at the micron scale. These structures can be miniaturized, remotely controlled, wirelessly operated, and rapidly fabricated, achieving fast and reversible switching within milliseconds\textsuperscript{\cite{zhang2024sonotransformers}}. Acoustic shape-morphing methods can overcome the aforementioned challenges and pave the way for advancements in shape morphing. However, the mechanisms, driving methods, and control techniques for acoustic micro-machines are still in the early stages of research, and a comprehensive theoretical framework has yet to be established. 

Here, we present acoustical shape-morphing micromachines(ASM) that can alter their shape within 10 milliseconds under acoustic field stimulation and rapidly return to their initial state once the acoustic field is removed. These micro-machines consist of a pair of identically sized microbubbles connected by flexible micro-hinges. When exposed to an acoustic field, the oscillation of the microbubbles induces deformation in the micro-machine within milliseconds. The forces and torques required for this deformation are provided by the secondary acoustic radiation forces between the microbubbles and the flow induced by their oscillation, and this deformation can be continuously tuned by varying the voltage of the exciting acoustic field. We also propose a modular design approach based on the rapid response, ultrafast deformation, large deformation angles, and programmability of these deformable micro-machines. Their motion can be predicted using Denavit-Hartenberg (DH) parameters, enabling deformation from 2D to 2D and 2D to 3D. Additionally, we demonstrate the potential of this method for 3D to 3D deformation. These results highlight the remarkable capabilities of micro-machines in terms of deformation modes, programmability, biomimetic design, and motion, with extensive applications at the microscopic scale, including soft robotics, flexible electronics, microfluidics, smart materials, and adaptive optics.

\section{Results}\label{sec2}

\subsection{Design and Characterization of ASMs}\label{subsec2}

Microbubbles are widely used acoustic tools at the microscale. Compared to other tools, such as sharp-edged structures\textsuperscript{\cite{harley2024enhanced}}, microbubbles exhibit frequency response characteristics under external acoustic field stimulation\textsuperscript{\cite{wei2024acoustic}}. Additionally, there are interaction forces between different microbubbles\textsuperscript{\cite{kaynak20233d}}. These mechanisms offer greater flexibility in the design and characterization of microbubble-based micromachines, enhancing their potential multifunctionality. The fabrication process is depicted in the figure (see supporting information, Figure1). We designed and manufactured micromachines with hinges of varying lengths, ranging from 4 to 16 µm, and thicknesses from 0.5 to 3 µm, but with a uniform height of 57.5 µm in the z-direction. The variable deformable angle between the two microbubbles is between 0 and 75 degrees. For more detailed information on the fabrication process, please refer to the Materials and Methods section.

We characterized these ASMs using an experimental setup that includes an acoustic manipulation region and a piezoelectric transducer (PZT) affixed adjacent to it, both mounted on a glass slide (see Supporting Information, Figure2). The piezoelectric transducer is connected to a signal generator through an amplifier to produce an acoustic field with adjustable amplitude and frequency. The dynamic deformation of the ASMs was observed using a microscope and a camera. We applied an acoustic field with a voltage amplitude ranging from 0 to 60 V and a working frequency between 102 and 104 kHz, which matched the resonance frequency of the experimental setup. More detailed information can be found in the Materials and Methods section.

A pair of microbubbles with identical diameters are crucial to the ASM, providing the deformation capability for the entire device. When the acoustic field passes through the liquid medium and reaches the micromachine, acoustic streaming effects occur between the pair of microbubbles, while secondary acoustic radiation forces induce the bubbles to attract each other. Simultaneously, the microhinges contract, causing the two microbubbles to approach each other until they make complete contact. The ASM can achieve a fully folded shape within 5 to 10 milliseconds (as indicated by experimental results). This dynamic deformation is controlled by acoustic pressure; once the acoustic field is turned off, the ASM rapidly returns to its original shape, as illustrated in Fig. 2(a) to Fig. 2(d). The dynamic deformation is regulated by the excitation voltage of the external acoustic field, which can be adjusted via the signal generator.

To further understand the mechanisms underlying shape-morphing, we investigated the dynamic mechanisms of the micromachines within the acoustic field. According to bubble acoustics, acoustically excited bubbles act as acoustic sources when they approach each other\textsuperscript{\cite{kaynak20233d,doinikov1995mutual}}. Since the distance between the bubbles is much smaller than the wavelength of the ultrasound in the experimental setup, it can be assumed that the acoustic radiation force generated by the background acoustic field does not cause relative displacement between the microbubbles. The secondary forces acting on the micromachine are therefore the interaction forces between the bubbles, resistance induced by acoustic streaming, and secondary acoustic radiation forces. Identical bubbles are expected to produce the same acoustic streaming, and thus, the interaction force between identical bubbles will be attractive, which causes adjacent microbubbles to push away from each other. Therefore, the total force acting on the microbubbles is:

\begin{equation}
    F_B=F_U+F_S+F_R
\end{equation}

where $F_U$ represents the acoustic radiation force, $F_S$ denotes the thrust from the flow, and $F_R$ refers to the drag force exerted on the bubble by the flow generated by neighboring bubbles. In our work, since the distance between bubbles is always comparable to the bubble size and the bubble size is significantly smaller than the wavelength of the sound wave, we assume that $F_R$ does not depend on the distance between the microbubbles\textsuperscript{\cite{bruus2012acoustofluidics}}. Therefore, we have $|F_R|=|F_S|$. According to the force balance equation\textsuperscript{\cite{doinikov2005bjerknes,doinikov1995mutual}}, the equation can be rewritten as

\begin{equation}
    F_B=A*G
\end{equation}

\begin{equation}
    G=\dfrac{r_0^2}{L^2}\dfrac{1}{[(1-\omega_0^2/\omega^2)^2+\delta_0^2]}
\end{equation}
 
where $G>0$ represents attraction. As shown in the figure, $r_0$ represents the radius of identical microbubble pairs, $L$ denotes the distance between the microbubbles, $\omega_0$ is the natural frequency of the microbubble pair, and $\omega$ is the acoustic excitation frequency. $\delta_0$ denotes their total damping constants. $A$ is a scalar related to the excitation amplitude and does not change the sign of the force. From Equation (3), it can be observed that the sign of the interaction force between identical microbubbles remains constant regardless of changes in the acoustic excitation frequency. Therefore, identical bubbles tend to attract each other. In contrast, the interaction force between microbubbles of different sizes is more complex, potentially resulting in either attraction or repulsion, depending on the acoustic excitation frequency.

We investigated the dynamics of the ASM using high-magnification microscopy and observed an intriguing phenomenon: as two microbubbles are driven by a variable amplitude acoustic field, the deformation amplitude of the micromachine changes accordingly. When the amplitude of the acoustic field is constant, the deformation amplitude of the micromachine also remains constant. By introducing tracer particles into the liquid, rapid oscillations of the microbubbles were observed, generating micro-vortices in the surrounding fluid (see Supporting Information, Figure3). As the distance between the two microbubbles decreases, according to Equation (3), their interaction force significantly increases with decreasing distance. We observed that the microbubbles progressively attract each other, inducing the folding of the microhinges. The angle between the two microbubbles decreases nonlinearly, as illustrated in the fig. 2(h).

We conducted a series of experiments to optimize the micro-hinge structure of ASMs by varying the length, angle, and thickness of the flexible micro-hinges with the goal of facilitating efficient transformation. Additionally, we observed that the deformation angle of the micromachine is significantly influenced by the thickness and length of the micro-hinges. Under a voltage excitation of 0-60 V, reducing the thickness of the micro-hinges by 50\% (from 2 µm to 1 µm) resulted in a 445\% increase (from 11° to 60°) in the maximum deformable angle at 10 µm, as shown in the figure. This can be attributed to the reduction in stiffness. Similarly, increasing the length by 67\% (from 6 µm to 10 µm) led to a substantial 100\% increase (from 30° to 60°) in the maximum deformable angle at a thickness of 1.5 µm. When the micro-hinge length does not exceed 10 µm, the acoustic radiation force dominates the total force. However, when the micro-hinge length exceeds 10 µm, the magnitude of the acoustic radiation force decreases, and the thrust from the fluid becomes dominant, causing the sign of the total force to switch, resulting in the microbubbles beginning to repel each other, as shown in the fig. 2(h) to fig. 2(k).

We also found that as the initial angle of the ASM increases, the excitation voltage required for complete deformation increases, as depicted in the figure. Furthermore, we investigated the deformation time of the micromachine under different thicknesses and excitation voltages. We found that when only varying the external excitation voltage, the micromachine requires 50-60 ms to deform fully if it has not reached complete deformation. Once full deformation occurs, the deformation time decreases significantly with increasing excitation voltage, reaching only 8 ms at 60 V. Consequently, thinner and longer micro-hinges have relatively lower stiffness. Therefore, soft hinges can generate greater acoustic forces, leading to more pronounced deformation. Moreover, by adjusting the laser intensity, the degree of cross-linking of the micro-hinges can be controlled, selectively enhancing the stiffness differential between the rigid link and the soft hinge. The wide range of options for modulating stiffness provides exciting new possibilities for designing and programming transformable micromachines and microrobots. Detailed theoretical development, validation, and simulations of the model are provided in the Supporting Information appendix.

\subsubsection{Design and Encoding of 2D ASMs}\label{subsubsec2}

Unlike traditional 3D printing, shape-morphing micromachines require addressing complex forward and inverse design problems, making the design of models crucial in experimentation. Considering the degrees of freedom and programmability of the ultrasonic deformation module, we employ a modular design approach to analyze the forward and inverse kinematics of the target structure by defining the Denavit-Hartenberg (DH) parameters associated with deformation. The DH parameters adhere to the principle of superposition, describing any complex transformation with only four physical parameters in a closed-form analytical manner\textsuperscript{\cite{ denavit1955kinematic}}.

\begin{equation}
    [T]=[XY_1][Z_1][XY_2][Z_2]\cdots[XY_n][Z_n]
\end{equation}

In this context, $T$ is the transformation that positions the terminal link of the last building module. This transformation allows the frame structure to undergo a curved motion around a defined axis.

\begin{equation}
    [XY_i]=Trans_{xy_i}(R_{x_i})Trans_{y_i}(d_i)Rot_{xy_i}(\theta_{xy_i})Rot_{xy_i}(\alpha_{xy_i})
\end{equation}

\begin{equation}
[Z_i]=Trans_{z_i}(d_i)Rot_{z_i}(\theta_{z_i})
\end{equation}

where $d_i$, $\alpha_i$, $R_i$, $\theta_i$ are referred to as the DH parameters, which in combination with the micro-machine units described above define the morphology planning principles of the modular system, as shown in fig.3 (a) to fig.3 (i). Fig. 3 (a) to fig. 3 (e) show the simulated rotational motion of micromachine units encoded using $\theta_i$. These microunits are rigidly connected to each other to avoid deformation interactions between the microunits and to ensure the validity of the superposition principle. The rotation angle of the microunits rotating around the Z-axis is determined by $\theta_{z_i}$ and can be adjusted by the exposure dose during the manufacturing process. Fig. 3 (f) and Fig. 3 (g) indicate that by connecting the deformed microunits to the fixed microunits, the radius of rotation ($R_i$) can be changed. Fig. 3 (h) represents the microunits offset along the Z-axis ($d_i$) to construct the deformed morphology. Fig. 3 (i) illustrates that a new coordinate system can be reconstructed by rotating the microcell in the XY plane ($\alpha_i$). Together, the above four parameters form the morphological planning rule for the modularized system.

According to the above rules, for any 2D shape, we can convert it to a finite number of ASM modular. Then, the DH parameters are obtained by inverse kinematics and switching rules are constructed between the target morphology and the initial morphology according to Eq. (4), as shown in fig. 4 (a) and fig. 4 in supporting information. For example, the deformation design of the chain as the target morphology and the wave as the initial morphology. Here, $\theta_i$ is the only variable parameter in the morphing design, and the specific parameter information is shown in Table 1 (Supporting Information). Fig. 4 (b) and fig. 4 (d) shows the experimental results of the conversion from chain to arc. Under the excitation voltage of 23V, the chain structure folds sequentially, and the deformation of four modular produces a deformation effect of up to 233 degrees, and it only takes 2003ms to complete the folding and 742ms to complete the release, as shown in fig. 4 (e). In the next experiments, we observed that the response time was longer as the voltage increased before the full folding was completed. And once the excitation voltage exceeds the critical voltage of 23V at which the full folding can be done, the time to complete the full folding gradually decreases until it takes only 95ms to deform from chain to arc at an excitation voltage of 50V. In addition, Fig. 4 (f) to (i) show the results of transforming from chains to rolls, “S ”s, honeycombs, and waves (see movies ()-()), demonstrating the versatility and generality of our design approach.

In addition, multi-module micromachines that can store information can be composed using modular unit assemblies, as shown in fig. 5. Multiple tailored configurations can be realized by encoding the target traits according to the reverse design method. To demonstrate this, we perform different information edits for the chain structure so that the ASM can store the corresponding information. Here, we encode the letters 'P' and 'U' in the alphabet into the chained 2D-ASM as shown in fig. (a). We characterize the deformation process of each letter, the dark-colored module is the driving part of the deformed structure, which gradually bends to the pre-edited position under the action of an externally stimulated acoustic field, as shown in fig. 5 (b) and fig. 5 (d) and video. Thus, this ASM concept of information coding allows the creation of complex shape coding systems with customized 2D transformations by customizing the design and layout of the deforming microunits.

\subsubsection{3D Complex ASM}\label{subsubsec3}

\textbf{Blooming micro-lotus flower} \par
To further extend the application scenarios of acoustic ASM technology, we considered the complex configurations of ASM in 3D space (called 3D-ASM). According to the aforementioned deformable micromodules, under varying intensities of acoustic fields, the interaction forces between bubbles enable them to maintain a stable state under any deformed configuration. We adjusted the deformation direction of the micromodules to extend the degree of deformation freedom along the Z-axis. Specifically, we designed an acoustically driven micro-lotus flower inspired by the blooming of lotus flowers in nature (see Figure 6(a)). To address the constraints between the structures, the flower comprises four petals, which are connected to the center by ASM micromodules, as illustrated in Figure 6(b). The driving force for the deformed structure is provided by the interaction forces between a pair of identical microbubbles activated by the acoustic field. By configuring the deformation direction of the micromodules, we enabled them to deform along the Z-axis, achieving a maximum deformation angle of 60 degrees. we illustrate the deformation process in chronological order using one petal of the flower as an example. Excitingly, compared to deformation methods such as pH\textsuperscript{\cite{chen2021intelligent}}, thermal\textsuperscript{\cite{kim2023shape}}, and photonic stimuli\textsuperscript{\cite{yang2019photoresponsive}}, The deformation of our micro-lotus flower can be halted at any moment and maintained in its current configuration by fixing the excitation voltage. In this state, when external disturbances act on the micro-lotus flower, it can quickly return to a state of equilibrium. We applied significant disturbances to the deformed micro-lotus flower using a micro-probe, and within () milliseconds, the shape of the micro-lotus flower was restored, as shown in Supplementary Figure (). When the ultrasonic excitation signal is halted, the flower rapidly returns to its initial state. This deformation method, which allows for rapid opening and closing while maintaining the current state at any moment, enables precise three-dimensional manipulation of microparticles, cells, or micro-robots at the microscale.

Furthermore, as shown in Figure 5(c), we observed that the deformation time of the micro-lotus flower gradually decreases with increasing voltage. The maximum time required for the structure to completely deform from the "bloom" state to the fully deformed "bud" state is 621 ms, when the external acoustic field is activated with a voltage of 30 V and a frequency of 103 kHz. As the excitation voltage continues to increase, the time required for complete deformation further decreases. When the acoustic field is activated with an excitation voltage of 50 V, the entire deformation process can be completed in 138 ms, resulting in a 77.8\% reduction in response time. As shown in Figure 6(e), we present the optical images of the micro-lotus flower deformation process under an excitation voltage of 40 V and an excitation frequency of 103 kHz. In addition, to better correlate the optical images with the degree of deformation, we obtained the projection curves of the micro-lotus flower in the XY plane at different deformation angles, as illustrated in Figure 6(d).

\textbf{Multi-flight-mode micro-bird} \par
More advanced deformation behaviors can be achieved by editing the deformation direction, while the adjustment of the deformation amplitude can be realized through the maximum deformation angle between the microbubbles within the ASM. In the example shown in Figure 7(c), two ASM units with different deformation directions are oriented along the Z-axis while being distributed at different orientations in the XY plane. By adjusting the deformation direction and amplitude of each ASM, arbitrary deformation poses can be assigned to each component. As a demonstration, we designed a micro-bird resembling origami to mimic the different flight modes of real birds (see Figure 7(a)). This is achieved through the specific arrangement of ASMs on five body parts of the micro-bird: the head, body, tail, and a pair of wings, with the corresponding electron microscopy image shown in Figure 7(b). At the wings or joints of the micro-bird, there are two possible deformation directions and one stationary state. By editing the configurations of the ASM on various parts of the micro-bird, we demonstrate four distinct morphological changes: "flapping," "taking off," "turning," and "hovering," as illustrated in Figure 7(d) and the supplementary video. Here, various deformation behaviors are exhibited: in the flapping mode, both wings fold simultaneously in phase and in the same direction; in the taking off mode, the head and tail rise while the wings fold downward; in the turning mode, the wings fold in different directions; and in the hovering mode, only the tail folds relative to the body. These results indicate that by using ASM as a foundational structural module, a wide range of micro-deformable micromachines can be designed, offering diverse options in terms of structural configurations, geometric dimensions, and deformation directions.

\section{Discussion}\label{sec12}

Acoustically deformable micromachines, particularly those utilizing materials with varying stiffness, remain largely unexplored. In our work, we demonstrate the ASM based on the interaction forces between microbubbles. These micromachine modules consist of a pair of microbubbles connected by flexible micro-hinges, allowing them to change shape within milliseconds and rapidly return to their original state upon cessation of the acoustic field. Our simulations and theoretical studies further validate this novel shape transformation mechanism. Meanwhile, we proposed a 2D programmable design method for ASM-based micromodules to facilitate the forward and inverse problems of complex deformable structures, along with an exploration of information storage. In more complex 3D-to-3D shape-morphing micromachines, our device demonstrates extraordinary reliability, stability, and flexibility. These shape-morphing micromachines are capable of withstanding external disturbances, such as the blooming micro-lotus, under acoustic field excitation of arbitrary amplitudes. Furthermore, they can be designed for specific design such as the micro bird with various flight modes. 

Acoustically multifunctional deformable micromachines pave the way for advancements across multiple scientific domains. In the field of microfluidics, integrating our micromachines into microchip platforms facilitates their application in drug delivery, object loading, and controlled screening. In the field of metamaterials, designing complex arrays of deformable microstructures can be utilized for purposes such as optical steering, acoustic guiding, and energy amplification. In the field of microrobotics, acoustically deformable micromachines provide a new design paradigm for acoustic microrobots, enhancing their maneuverability and environmental adaptability through functionalities such as motion and navigation. For flexible electronic devices, deformable flexible micromachines can facilitate enhanced interaction, information transmission, and functional development. Furthermore, in the biomedical field, due to the high compatibility of acoustic fields with biological tissues, these micromachines may be applied in spaces such as blood vessels and the gastrointestinal tract in the near future.

\bmhead{Acknowledgements}

We acknowledge financial support from the Key Projects in Shaanxi Province (No.2024CY2-GJHX-08), the Key Projects in Shaanxi Province (No.2024CY2-GJHX-10) and China Postdoctoral Science Foundation (No.2023M740091)

\bibliography{sn-bibliography}% common bib file

\section{Figures}\label{sec6}

\begin{figure}[h]
\centering
\includegraphics[width=0.9\textwidth]{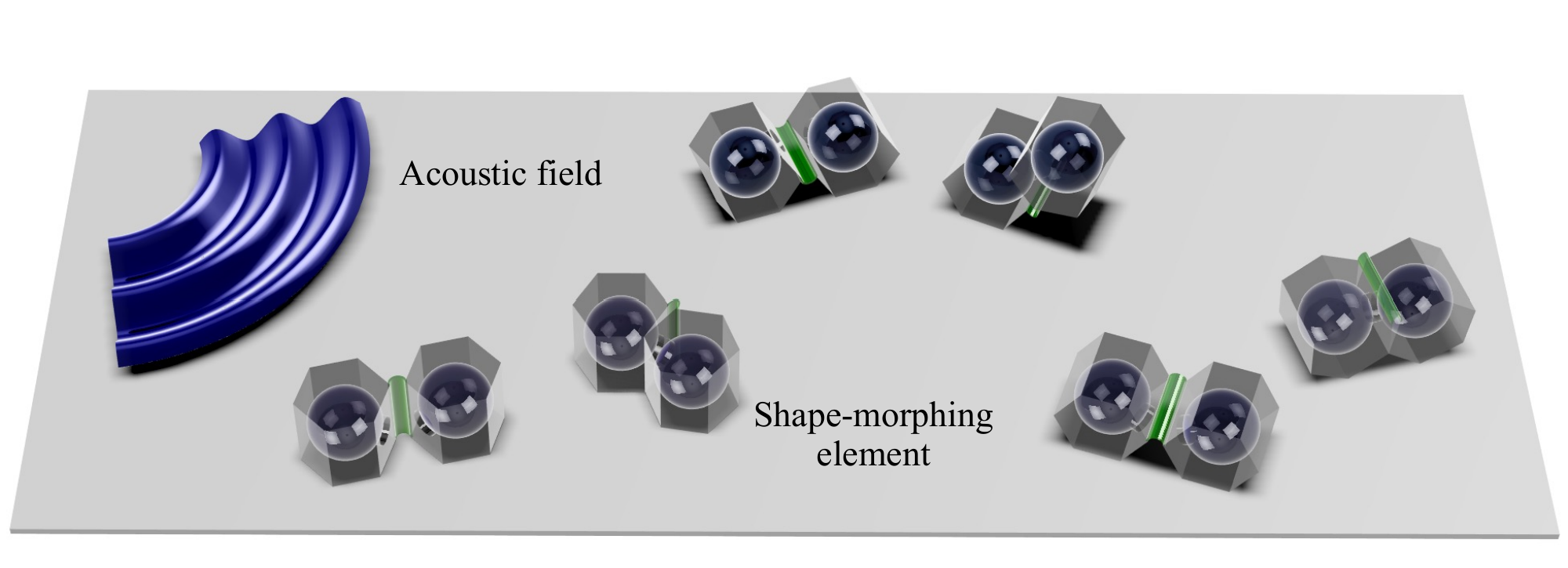}
\caption{Morphing behavior of acoustic shape-morphing micromachine. (a) Schematic of the micromachine morphing due to acoustic streaming effect. Micromachine exhibits a certain morpging mode selected by the sound wave of a certain excitation intensity and frequency.}\label{fig1}
\end{figure}

\begin{figure}[h]
\centering
\includegraphics[width=0.9\textwidth]{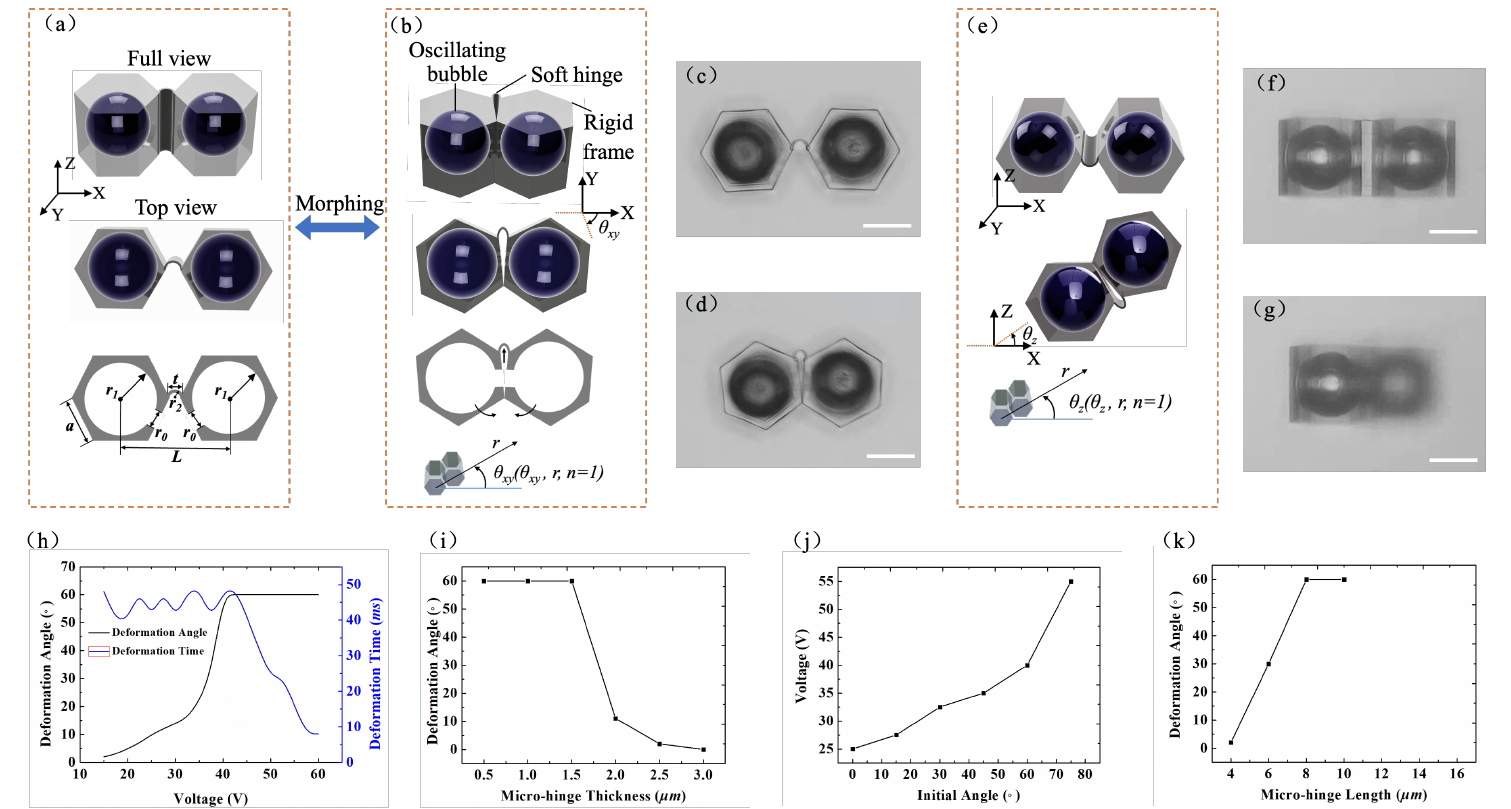}
\caption{Characterization and optimization of the fabricated ASM. (a) Schematic and design geometry of an ASM in the xy-plane, whose basic structure is a pair of microbubbles of the same diameter and a flexible micro-hinge connecting them. (b) Schematic of the ASM in the xy-plane after deformation under acoustic field excitation, with two microbubbles in close proximity to each other while the flexible micro-hinge bends. (c)-(d) Optical images of the initial and deformed ASM in the xy plane. (e) Schematic and design geometry of an ASM in the zx-plane. (f)-(g) Optical images of the initial and deformed ASM in the zx plane. (h) Effect of acoustic field excitation voltage on ASM deformation angle and deformation time. (i) Effect of micro-hinge thickness on ASM deformation angle. (j) Relationship between the initial angle of the ASM and the excitation voltage required for its complete deformation. (k) Effect of micro-hinge length on ASM deformation angle.}\label{fig2}
\end{figure}

\begin{figure}[h]
\centering
\includegraphics[width=0.9\textwidth]{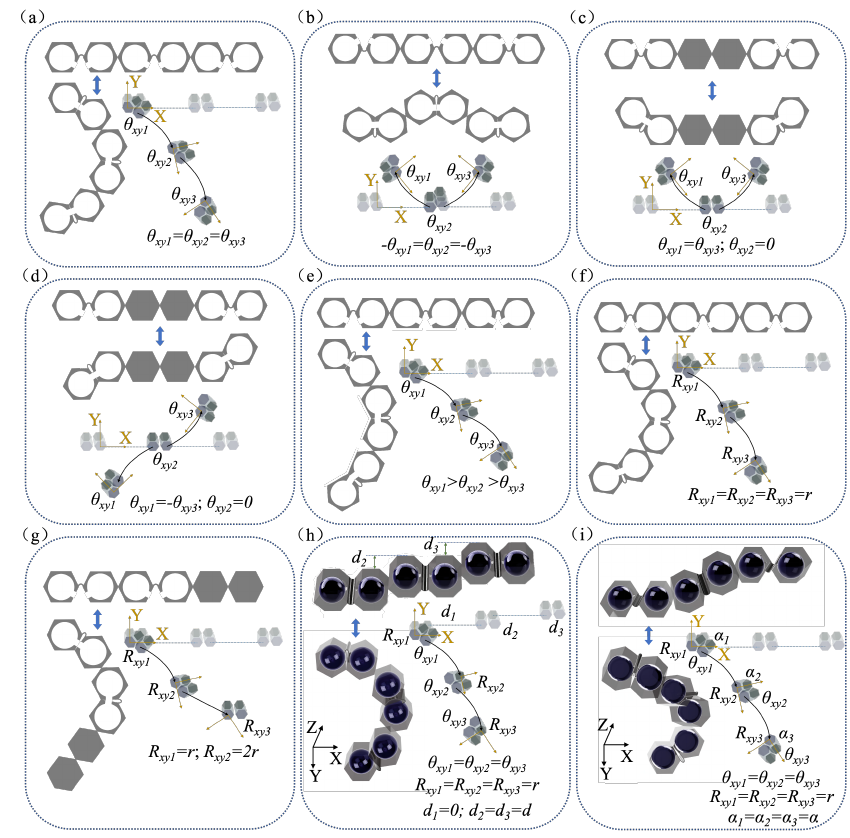}
\caption{Design principles and rules for ASM modular systems. (a)-(e) Rotational deformation induced by the morphing of the ASM. Each ASM can be viewed as a combination of a rotational joint and a rigid bar, resembling a robotic arm. (f)-(i) Schematics of rotational movements with controlled amplitude and orientation enabled by the assembly of various preprogrammed ASM. (a)-(i) define how the four DH parameters $\theta$, R, d, and $\alpha$ are implemented in our modular ASM, respectively.}\label{fig3}
\end{figure}

\begin{figure}[h]
\centering
\includegraphics[width=0.9\textwidth]{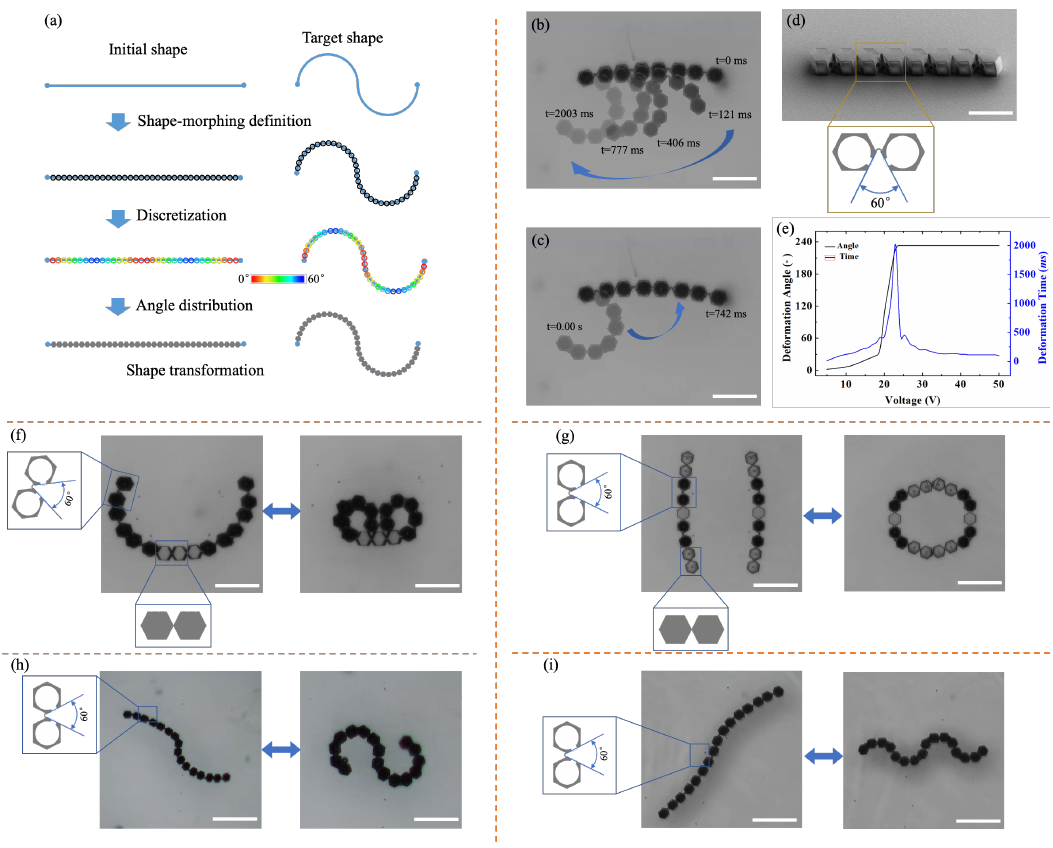}
\caption{Inverse and forward design of morphing modular systems. (a) Inverse problem finding for programming a structure that morphs into the desired shape. For an arbitrary shape, such as a wave, the modular design converts it to a discrete counterpart with a finite number of joints and then obtains the DH parameters. The modular system subsequently constructs the shape transformation between the given wave shape and an assembled roll configuration by encoding the inversed $\theta_z$ into the roll, for it to morph into the shape of a wave. (b) The chain structure morph into an arc under sound field excitation, and the maximum time required for the ASM to complete the full deformation is 2003 ms. (c) After stopping the acoustic field excitation, the arc structure returned to the chain structure after 742 ms. (d) SEM images of the ASM. (e) Relationship between ASM deformation time and angle and sound field excitation voltage. (f)-(i) Optical images of the assembled morphing ASM encoded with different DH parameters. }\label{fig4}
\end{figure}

\begin{figure}[h]
\centering
\includegraphics[width=0.9\textwidth]{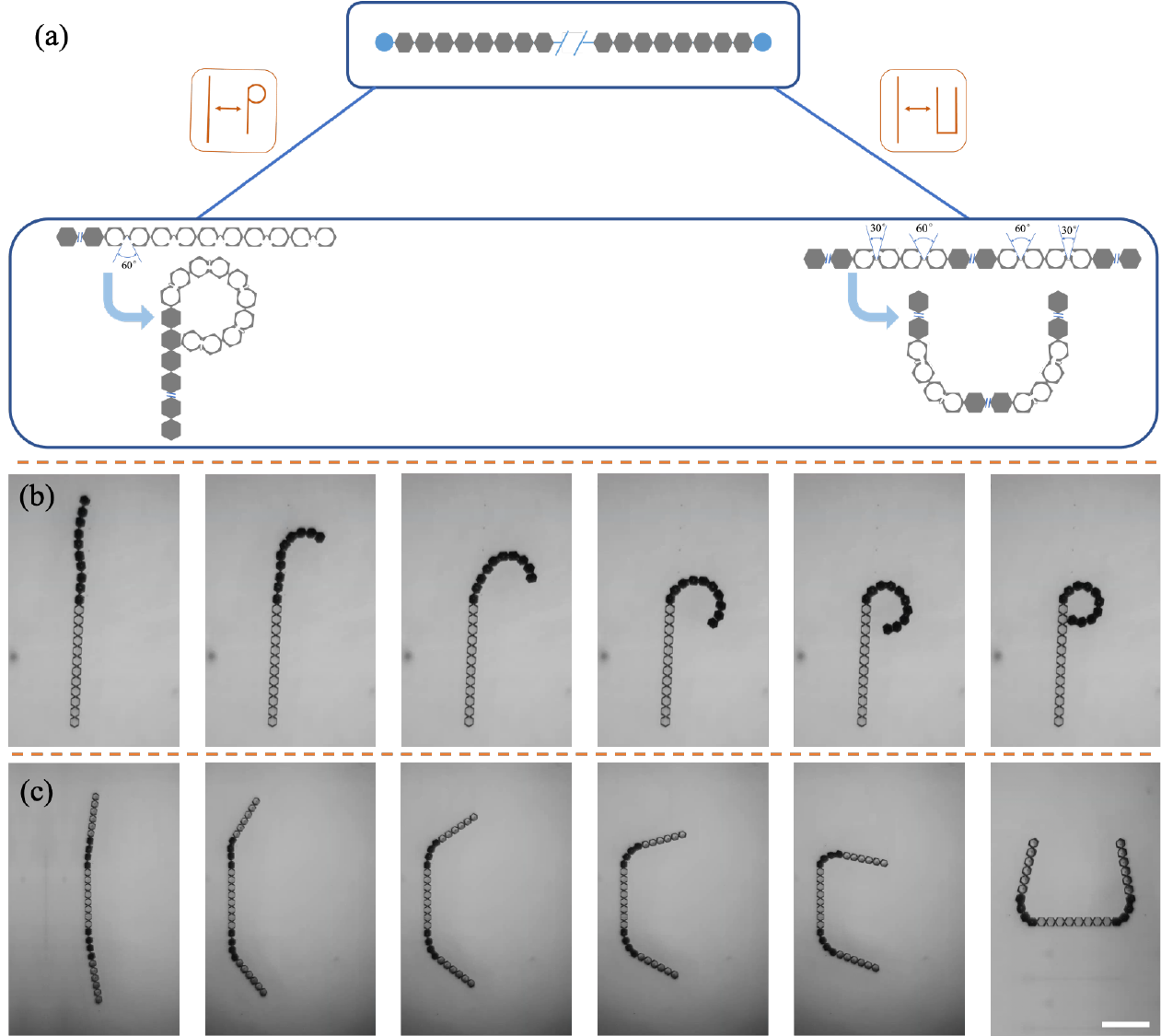}
\caption{Encoding letters of the alphabet into ASM. (a) Conceptual design of ASM with chain structure that can be encoded to transform into letters ‘P’ and ‘U’. Each module in the schematic can be coded independently and the deformation angle can be selected independently. As a result, the characteristic positions of the chain structure will be deformed or fixed when a sound field is applied. (b)-(c) Schematics and corresponding optical microscope images of the fabricated micromachines encoded for ‘P’ and ‘U’ shape morphing. The corresponding voltage of the excitation sound field is 30V. }\label{fig5}
\end{figure}

\begin{figure}[h]
\centering
\includegraphics[width=0.9\textwidth]{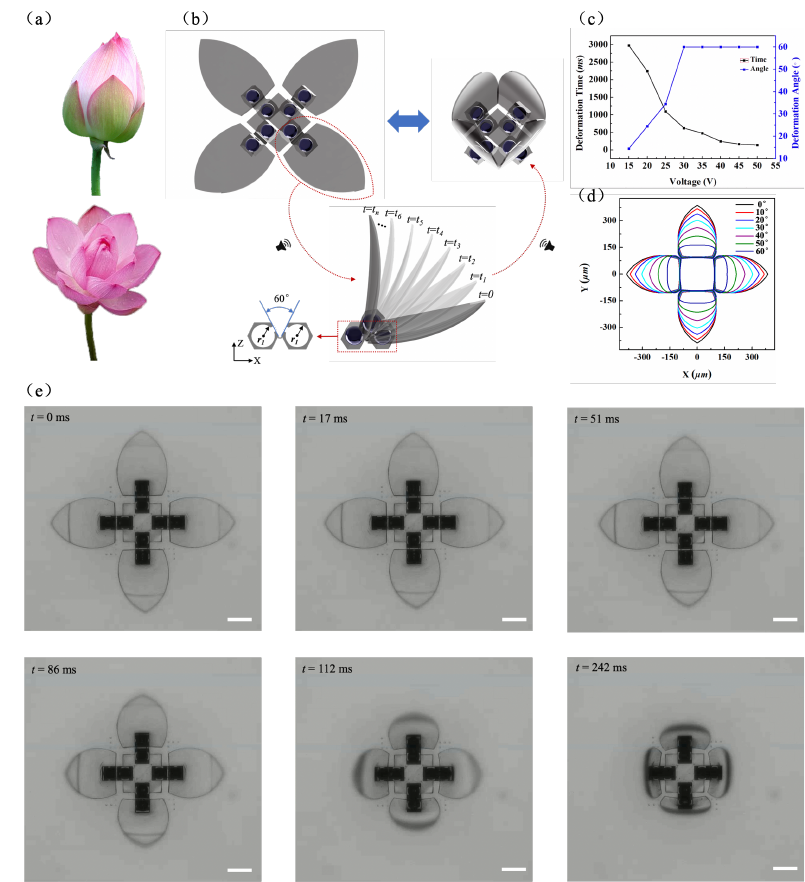}
\caption{3D Lotus-ASM with reversible transformation and infinite-mode. (a) Shape-morphing occurs across in nature represented by a lotus flower. (b) Schematic of an untethered acoustic micromachine and its high-speed, reversible shape transformation in an acoustic field. (c) Effect of acoustic field excitation voltage on the 3D Lotus-ASM deformation angle and deformation time. (d) The 3D Lotus-ASM projection curves on the xy plane for different deformation angles. (e) The 3D Lotus-ASM  constructed from 'bloom state' folds into 'budding state' with an excitation voltage of 40V for the external acoustic field. }\label{fig6}
\end{figure}

\begin{figure}[h]
\centering
\includegraphics[width=0.9\textwidth]{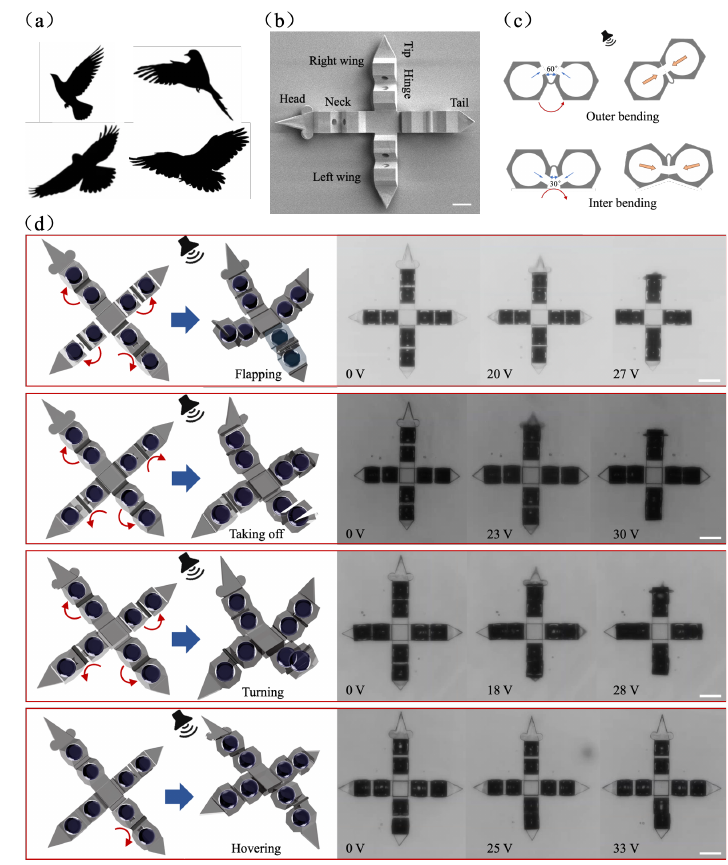}
\caption{Origami-like 3D Bird-ASM with multiple shape-morphing modes. (a) The multiple flying postures of birds in nature. (b) SEM image of the 3D Bird-ASM. (c) Schematic of the morphing behaviours of 3dD ASM when actuated using an applied acoustic field. (d) Schematic (left) and optical images (right) of a 3D Bird-ASM mimicking four flying modes, ‘flapping’, ‘taking off’, ‘turning’ and ‘hovering’. From left to right in each row: schematic of a flat 3D Bird-ASM with each direction of morphing in 3D ASM; schematic showing the folding of the 3D Bird-ASM under the acoustic field, and the optical microscope images of the experimental demonstrations. For each flying mode, the three optical images show the shape-morphing in acoustic field with different intensity.}\label{fig7}
\end{figure}

\end{document}